\documentclass{emulateapj}

\usepackage{amsmath}
\usepackage{graphicx}
\usepackage{natbib}

\newcommand{\hii}{H~{\sc ii}}

\begin{document}

\title{Are molecular outflows around high-mass stars driven by ionization feedback?}

\author{Thomas Peters\altaffilmark{1,2,3}}
\email{tpeters@physik.uzh.ch}

\author{Pamela D. Klaassen\altaffilmark{4,5}}

\author{Mordecai-Mark Mac Low\altaffilmark{6}}

\author{Ralf S. Klessen\altaffilmark{1}}

\author{Robi Banerjee\altaffilmark{7}}

\altaffiltext{1}{Universit\"{a}t Heidelberg, Zentrum f\"{u}r Astronomie,
Institut f\"{u}r Theoretische Astrophysik, Albert-Ueberle-Str. 2,
D-69120 Heidelberg, Germany}
\altaffiltext{2}{Fellow of the Baden-W\"{u}rttemberg Stiftung}
\altaffiltext{3}{Institut f\"{u}r Theoretische Physik, Universit\"{a}t Z\"{u}rich, Winterthurerstrasse 190, CH-8057 Z\"{u}rich, Switzerland}
\altaffiltext{4}{European Southern Observatory, Karl-Schwarzschild-Strasse 2, Garching, Germany}
\altaffiltext{5}{Leiden Observatory, Leiden University, PO Box 9513, 2300 RA Leiden, The Netherlands}
\altaffiltext{6}{Department of Astrophysics, American Museum of Natural History,
79th Street at Central Park West, New York, New York 10024-5192, USA}
\altaffiltext{7}{Hamburger Sternwarte, Gojenbergsweg 112, D-21029 Hamburg, Germany}

\begin{abstract}
The formation of massive stars exceeding 10 $M_\odot$ usually results in large-scale molecular outflows.
Numerical simulations,
including ionization, of the formation of such stars show evidence for ionization-driven molecular outflows. We here examine whether the outflows seen in
these models reproduce the observations. We compute synthetic ALMA and CARMA maps of CO emission lines of the outflows, and
compare their signatures to existing single-dish and interferometric data. We find that the ionization-driven models can
only reproduce weak outflows around high-mass star-forming regions. We argue that expanding \hii\ regions probably do not represent the
dominant mechanism for driving observed outflows. We suggest instead that observed outflows are driven by the collective action of the outflows
from the many lower-mass stars that inevitably form around young massive stars in a cluster.
\end{abstract}

\maketitle

\section{Introduction}
\label{sec:intro}

Jets and outflows always accompany the star formation process, both in the low-mass and in the
high-mass regime \citep{caband91,bachiller96,reibal01,shepherd05,beuthshep05,arceetal07,baletal07,bally07,bally08}.
Though outflows from low-mass stars are in general understood as resulting from magnetically-driven disk winds during the accretion process, the formation of high-mass outflows from
ionizing stars with masses in excess of $10\,M_\odot$
is less understood, as we argue in the next Section, both from the observational (Section~\ref{sec:obsout})
as well as the theoretical (Section~\ref{sec:theo}) point of view.

The recent finding by \citet{petersetal10a} that ionization feedback from massive protostars
can drive bipolar molecular outflows has suggested that these could explain the observations of high-mass outflows.
In the present paper, we compare the morphology of these molecular outflows
with synthetic CO $J = 2 - 1$ observations (Section~\ref{sec:synobs}).
We demonstrate that the synthetic observations fail to consistently reproduce 
observations of molecular outflows in Section~\ref{sec:disc}. In Section~\ref{sec:conc},
we discuss the implications and limitations of this result, and speculate on the
actual origin of the observed outflows.

\section{Observations and Theory of High-Mass Outflows}

\subsection{Observations of High-Mass Outflows}
\label{sec:obsout}

Bipolar outflows from high-mass stars are, in many ways, scaled-up versions of their lower-mass
counterparts. They do, however, carry significantly more momentum \citep[e.g.][]{beutheretal02,wuetal04},
and are often less collimated \citep[e.g.][]{beuthshep05}.

Outflows are launched from accretion disks around young stellar objects \citep[e.g.][]{rayetal07}. Disk-like structures
have recently also been found around high-mass protostars \citep{chini04,chini06,beltranetal04,
beltranetal06,jiangetal08,daviesetal10,zapataetal10,motogietal11}, and even in situations where disks cannot be resolved,
rotation of the molecular \citep{howardetal97,zhangetal98,klaassetal09,gavmadetal09,beltetal11a,furuyaetal08,furuyaetal11}
and ionized \citep{garayetal86,sewiloetal08,ketoklaas08} gas can still be detected. In many of these
regions, bipolar molecular outflows originating from the center of rotation and oriented perpendicular to the rotation have been detected as well,
suggesting a causal connection between rotation and outflow.

If high-mass outflows were driven magnetically like their lower-mass analogs, then one would
expect a corresponding magnetic field structure along the outflow. Indeed, recent observations
suggest a correlation between magnetic field orientation and the presence of outflows from
high-mass sources \citep{girartetal09,tangetal09a,tangetal09b,tangetal10,vlemmingsetal10,beutheretal10,surcisetal11}.
However, whether these maser and dust polarization measurements indicate that the
outflows organize the magnetic field or the magnetic field collimates the outflows is less clear.

The momentum and energy imparted to molecular outflows by high-mass protostars scales with the
luminosity of the powering source \citep{richeretal00,beutheretal02,arceetal07,lopsepetal09}. Once the high-mass protostar
becomes luminous enough to ionize its surroundings, the interaction between outflow and
\hii\ region further complicates the picture. Observations suggest that the outflows originate
within the ionized gas (\citealt{depreeetal94,gavmadetal09}, Klaassen et al.\ in prep.), and that they
can continue into a molecular outflow beyond the ionization boundary (e.g.~K3-50A; \citealt{depreeetal94}
and \citealt{howardetal97}). 

In order to make statistical arguments about the nature of outflows  
from protostars of at least intermediate mass, we rely on single dish  
surveys to characterize their properties.
\citet{shepchur96} looked at ten sources with bolometric luminosities
exceeding $10^2$ $L_\odot$ and only found 5 bipolar outflows at a
resolution of 60$''$, noting that for the other sources, two had multiple velocity components, and  
the other three had signal to noise ratios too low to draw conclusions from.
\citet{lopsepetal09}, at a spatial resolution of 11$''$, looked at higher luminosity
sources ($L_{\rm BOL} > 2\times10^4$ $L_\odot$) and found outflows for each of the nine sources they mapped,
and strong line wings in the other two sources. Most of these last sources are 
associated with \hii\ regions.
\citet{qinetal08} studied the gas motions from 15 high-mass star forming regions with
ultracompact \hii\ regions and found evidence for outflows towards 10 of them. However, at a spatial
resolution $>1'$, this detection fraction should be used as a lower limit.
The five sources without outflows were amongst the nearest sources in their survey, suggesting
that distance did not play a role in the non-detections.
In each of these studies, outflow masses range from 3 to $10^3$ $M_\odot$, and their energies are of the
order $10^{46}$ erg or greater.

With interferometers one gains a better understanding of which
source is responsible for the large-scale outflows seen in the
single dish surveys.
With current instruments we are limited in the number of sources  
for which detailed analyses can be performed.
Nonetheless, in-depth observations have been made by many  
groups in an attempt to better understand the powering sources and  
their relationships with their environments.
Generally, these high resolution ($\lesssim 1''$) studies show that  
the outflow originates at the brightest continuum object in the  
field.  These sources include: G29.96 \citep{beltetal11a}, G24.78A \citep{furuyaetal02},
W51e2 \citep{ketoklaas08,shietal10}, G5.89 \citep{acordetal97,watsonetal07,hunteretal08},
NGC 7538 IRS 1 and G28.2 \citep{klaassenetal11}, G31.41 \citep{olmietal96b,olmietal96a},
G75.78 and G75.77 \citep{riflued10}.

In many of these sources, rotation of the warm gas surrounding the \hii\ region has also been detected  
perpendicular to the outflow direction (see \citealt{beltranetal06,beltetal11a,klaassetal09,olmietal96a,olmietal96b}, etc.).

\citet{ketoklaas08} and \citet{klaassenetal11} showed that  
molecular outflows appear at the edges of the \hii\ regions already at  
high velocities, suggesting these outflows are powered from within the  
\hii\ region. In fact, the radio recombination line study of \citet{depreeetal94}
displayed velocity shifts in this ionized gas tracer from K3-50A,
suggesting an ionized outflow. The molecular observations of  
\citet{howardetal97} are consistent with this outflow penetrating the  
ionization boundary and continuing as a molecular outflow.
Observations of ionized outflow lobes have also been reported for G48.75
\citep{johnstonetal09}.
Furthermore, \citet{cesaronietal10} discuss the possibility that
the outflow in G31.41 originates from an ionized jet, and \citet{guzmanetal11}
suggest that an ionized jet in IRAS 16562-3959 is the energy source of a molecular
outflow in the same region.

\subsection{Possible Outflow Driving Mechanisms}
\label{sec:theo}

We now consider the problems with existing explanations of outflow driving.
Protostellar outflows around low-mass protostars are generally attributed
to magnetic fields, either through the launching of disk winds \citep{blandfordpayne82,pudrnorm83,ouypud97,krasnoetal99,koniglpudritz00,fencem02,fendt06},
magnetic tower flows \citep{tomisaka98,tomisaka02,lynbel03,matstomi04,machidaetal04,banerjee06b,banerjee07,seifried12}
or through the interaction between the protostellar magnetosphere and the disk
field \citep{najitashu94,shuetal94,shuetal07,caietal08,lovelaceetal99,romanovaetal09}.
Recently, the applicability of these driving mechanisms in the high-mass
regime has been questioned \citep{petersetal11a} on the grounds that severe
gravitational instability in the accretion flow surrounding the massive star
destroys the coherent rotational motion in the disk plane that is required for all these
models to work, and the ionization feedback disrupts the
magnetic field structure. The statistical distribution of orbital separations with a peak at a semimajor axis
of $\sim 0.15\,$AU found by \citet{kobfry07} in the massive binary data of \citet{garmetal80} suggests that the inner disk region
is gravitationally unstable as well, making magnetic jet launching very difficult.
Furthermore, magnetically driven jets around low-mass stars may be collimated at the
disk-magnetosphere boundary \citep{liietal11}, and massive stars are known to have
relatively weak magnetospheres compared to low-mass stars \citep[e.g.][]{owocki09}.
To fully model this process we need to resolve the inner disk region (radii less than 1~AU)
where magnetic launching is expected. This will require high-resolution radiation-magnetohydrodynamical
calculations that connect the larger-scale gravitationally unstable accretion flow (radii of several
thousand AU) to the inner disk region.

\citet{vaidyaetal09,vaidyaetal11} tried to fill this gap in two-dimensional calculations that include both
ideal magnetohydrodynamics as well as the effects of radiative pressure in the line-driving approximation.
They find that initially well collimated outflows can widen up as the central star grows in mass and its
luminosity increases. This could explain the observational proposal that more evolved outflows around
massive stars have larger opening angles than very young ones \citep{beuthshep05}.
Because of the restricted geometry in the Vaidya et al. work, however, these studies cannot consider the effects of disk fragmentation,
so further analysis is required. A more consistent treatment of the complex interplay between the fragmentation
behavior of the inner disk, magnetic fields, and radiative processes has been presented by \citet{comm11}.
Their three-dimensional radiation-magnetohydrodynamics calculations indicate that magnetic braking
efficiently removes angular momentum from the disk, leading to high accretion rates and luminosities.
This heats the disk and reduces the degree of fragmentation at small radii. Because these calculations are
prohibitively expensive, they could not follow the evolution into the realm of massive stars where ionizing
radiation becomes important. This regime has been covered by \citet{petersetal11a} with somewhat lower resolution.
They find that magnetic fields cannot completely suppress fragmentation at large disk radii \citep[see also][]{hennetal11}.

Additional evidence against magnetic driving mechanisms comes from the notable lack
of observations of extremely high-velocity jets associated with high-mass outflows. Since
magnetically launched jets form deep within the gravitational potential well around the star,
and the potential well around high-mass stars is deeper than around low-mass stars,
jets would be expected with velocities comparable to those of line-driven stellar winds, exceeding 1000\,km\,s$^{-1}$.
Instead, the observed structures never propagate faster
than jets from low-mass stars. However, even if there were very
high-velocity jets, they might be immediately slowed by interaction with the accretion flow,
which is expected to be stronger for high-mass than for low-mass stars.

Radiation pressure on the accretion flow could also drive outflows.
\citet{krumholzetal09} used a gray flux-limited diffusion
method to simulate this effect. They
found that the radiation pressure creates bubbles around the massive star that
get destroyed by Rayleigh-Tayor instabilities. These radiation bubbles are
morphologically dissimilar to molecular outflows. Additionally, by the time
the bubbles form, the star would be massive enough to ionize the interior
of the bubble, so that the bubble would become an \hii\ region and not be
molecular anymore. \citet{kuiperetal11}, using a frequency-dependent treatment
of the direct stellar radiation combined with gray flux-limited diffusion,
do not see these Rayleigh-Tayor instabilities. As possible reasons for this
difference they suggest a better modeling of the first absorption of photons
with their numerical scheme, but it could also be that they will see the instability
once their bubbles reach the same size as those of \citet{krumholzetal09} or that
their density distribution has led to a decelerating, thus stable, bubble.
A more recent study \citep{kuiperetal12}, in which both the frequency-dependent treatment
and gray flux-limited diffusion are applied to the same initial conditions, seems to confirm
that the instability is an artifact caused by the gray approximation of the direct stellar radiation.
In any case, ionization would also fill the stable radiation-driven bubbles of \citet{kuiperetal11,kuiperetal12}
with ionized gas, so that stable radiation pressure-driven outflows cannot account for the observed high-mass
molecular outflows either.

Another source of momentum is line-driven winds
of high-mass stars \citep[e.g.][]{castoretal75}. Indeed, some of the observations
show high-mass outflows that are poorly collimated and seem to be consistent
with being wind-blown bubbles \citep[e.g.][]{shepherd05,beuthshep05}. However,
typical massive outflows have momentum fluxes
that are two orders of magnitude higher than those produced by stellar winds
\citep[e.g.][]{richeretal00}. Hence, stellar winds cannot be the primary
drivers of high-mass outflows already for reasons of momentum conservation,
and they are certainly not responsible for the more collimated
outflows found in earlier phases of high-mass star formation \citep[e.g.][]{beutheretal02}.

Recently, \citet{petersetal10a} have shown that ionization feedback can
drive bipolar molecular outflows. Contrary to classical understanding of \hii\ region
growth, these simulations show that the \hii\ regions around massive
protostars do not expand monotonically as long as the high-mass star is still
embedded in an accretion flow. Instead, the \hii\ region flickers whenever the ionizing
radiation gets shielded by dense filaments
formed in the gravitationally unstable accretion flow \citep{galvmadetal11}.
This rapid fluctuation of the thermal gas pressure drives
bipolar molecular outflows parallel to the net angular momentum vector: the pressure increase during
the ionization process drives the material away from the density enhancement in the plane of rotation, and the subsequent recombination
lets the blown-out material become a molecular outflow.

\section{Synthetic ALMA observations}
\label{sec:synobs}

We use synthetic ALMA observations of CO to characterize the ionization-driven outflows
found by \citet{petersetal10a}. We describe the input models (Section~\ref{sec:inpmod})
and the CO emission maps (Section~\ref{sec:coemma}).

\subsection{Input Models}
\label{sec:inpmod}

The simulations
use an extended version of the adaptive-mesh numerical hydrodynamics code
FLASH \citep{fryxell00} with sink particles \citep{federrathetal10} to follow the collapse beyond the formation
of the first objects, allowing us to describe the early evolution of the disk that builds up,
and our improved hybrid characteristics raytracing method \citep{rijk06,petersetal10a} that
propagates the ionizing and non-ionizing radiation on the grid.
We refer to \citet{petersetal10a} for details on the numerical algorithm and initial conditions.
A thorough description of the spatial and temporal \hii\ region evolution in
these simulations can be found in \citet{petersetal10a}, \citet{petersetal10b}
and \citet{galvmadetal11}.

We analyze two simulations. Both simulations start from a $1000\,M_\odot$, spherical molecular
cloud of diameter $3.2$\,pc. The cloud is initially in solid body rotation with
angular velocity $\omega = 1.5 \times 10^{-14}\,$s$^{-1}$. During the collapse
of this massive core, a rotationally flattened structure forms that quickly
becomes gravitationally unstable and fragments. In one simulation (Run~B),
we form a full stellar cluster in this rotating accretion flow, including radiative
feedback by both ionizing and non-ionizing radiation. In another simulation
(Run~A), also including radiation feedback, we artificially prevent the formation
of all stars after the first one and run the simulation with a single star only.
In Run~B, three massive stars with masses around $20\,$$M_{\odot}$ form, each creating
an \hii\ region around itself,
along with numerous lower mass stars. In contrast, the star in Run~A reaches more
than $70\,$$M_{\odot}$. A thorough discussion of the effects of radiative
feedback and the suppression of fragmentation can be found in
\citet{petersetal10a,petersetal10c}.

The massive stars in both simulations drive bipolar outflows
with their ionization feedback. The basic mechanism, as already mentioned
in Section~\ref{sec:theo}, is the repeated shielding of the ionizing radiation
by dense streams of gas in the accretion flow around the massive star.
The ionizing radiation accelerates the molecular gas around the \hii\ region down
the steepest density gradient, perpendicular to the rotational flattening.
When the ionizing radiation gets shielded by the filamentary accretion flow, the ionized gas can recombine
on timescales as short as $\sim 100$\,yr, and then the outflow becomes molecular,
given that the formation timescale of molecular hydrogen in these dense outflows is less
than a year \citep{holletal71}.
The molecular outflow continues to expand by momentum conservation. Since the outflow
is not driven anymore without the thermal pressure of the ionized gas,
the outflow can escape from the gravitational potential well of the
stellar cluster only if its initial velocity permits. After the loss of thermal support, the outflow
decelerates and eventually falls back onto the rotationally flattened structure by its gravitational attraction.
The online material of \citet{petersetal10a} contains movies of density slices showing this process for both runs.

Generally, the outflows in Run~A are much more powerful than the outflows
in Run~B. The reasons are the stronger accretion flow in which the massive star
is embedded and the larger ionizing luminosity. Though Run~A is not a realistic
model of massive star formation globally, the driving of the molecular outflow by the interaction of the
ionizing radiation with the accretion flow is modeled correctly. These stronger
accretion flows could be realized in simulations with multiple stars by,
for example, starting from a more massive core, using a more centrally concentrated
density profile \citep[e.g.][]{girietal12}, or reducing the initial rotational motion of the core \citep[e.g.][]{seifetal11}.

The online material for this article contains animations of the ionization-driven
outflows for both simulations. The videos show volume renderings of gas density (left)
and thermal pressure (right). The direct comparison of the two images allows the
identification of the \hii\ regions and illustrates how the thermal pressure of the
ionized gas drives the molecular fountains. For comparison, we also show an animation
of a third simulation, Run~E (see \citealt{petersetal11a}), in which an additional large-scale
outflow is driven by magnetic fields and interferes with the ionization-driven outflows.
Though the interplay between the different types of outflows is highly complex,
the morphological differences between the magnetically-driven outflows and the ionization-driven
outflows becomes very clear.

\subsection{CO Emission Maps}
\label{sec:coemma}

We have previously used the three-dimensional adaptive-mesh radiative transfer code
RADMC-3D\footnote{http://www.ita.uni-heidelberg.de/$\sim$dullemond/software/radmc-3d/}
to generate maps of free-free and dust
emission from our simulation data \citep{petersetal10b}.
Here we use the capability of RADMC-3D to model molecular line emission
\citep[e.g.][]{shettyetal11}. In order to trace the bulk dynamics of the gas,
we have chosen to model the CO $J=2-1$ transition at 230.538~GHz.
We have simulated observing the outflows at a distance of 1.3~kpc, the same distance as G5.89-0.39 \citep{motogietal11}. This source
is believed to be powered by a 25~$M_\odot$ star, and is so directly comparable with one of the stars in Run~B.

CO is an excellent tracer of the bulk gas dynamcis in star-forming regions because of its
relatively low critical density, and relatively high abundance ($10^{-4}$ with respect to molecular hydrogen, \citealt{wilroo94}).
It is the most abundant molecule after H$_2$ and is therefore a good tracer of large-scale structures.
Its low critical density means that it is detected even in the tenuous high-velocity gas in outflows.
SiO, another excellent outflow tracer, was not used because it emits on much smaller scales and its emission is
directly tied to the type of shocks produced in the outflow (i.e. Si is liberated from dust grains by sputtering during
the extended passage through C-shocks mediated by ion-neutral drag, but not by classical J-shocks \citep{schilkeetal97,gusdorfetal08,guilletetal11}).

For simplicity, we assume local thermodynamic equilibrium (LTE)
conditions. We use the CO Einstein coefficients from the Leiden
Atomic and Molecular Database \citep{schoeieretal05}.
These 3D data cubes were then converted into skymaps, and observations with ALMA were simulated using
the ``simobserve'' and ``simanalyze'' tasks in CASA 3.4 \citep{mcmullin07}. The native resolution of the original datacube was 49~AU.
Assuming the source was at the distance of 1.3~kpc, this gives a native resolution of $0.038"$.
This is much higher than the final resolution achieved in our ALMA simulations. We simulated observing with ALMA in a moderately high
resolution mode (``alma.out10.cfg''), achieving a resolution of 0.5$"$ at 230~GHz with a channel spacing of 1~km s$^{-1}$.
The rms sensitivity for the ALMA observations should be 1.2 mJy/beam for a 0.5$"$ beam.
Noise was added in the UV plane to simulate 4$^{\rm th}$ quartile observing conditions. We also simulated observing with CARMA in its
C configuration. This resulted in a final image resolution of $1.3"\times1.0"$. The CASA simulator assumed a system temperature of
200 K for the CARMA observations. For all simulations, we assumed a total integration time of four hours. For the ALMA mosaic, we
required six pointings, and with CARMA, two. 

Cleaning was done non-interactively using natural weighting and a 3 $\sigma$ cutoff. Zeroth and first moment maps
were created using line wing emission, excluding emission from within 3~km~s$^{-1}$ of the source rest velocity.
The zeroth moment (integrated intensity) maps were used to compute the column density following the procedure described in \citet{klaassenetal11}, assuming a temperature of $T=100\,$K.
Because the size of the emitting region can be determined
from the images, the number of molecules was calculated from the column density assuming a CO abundance of 10$^{-4}$ with
respect to H$_2$. From this, the gas mass was determined (see the first column of Table~\ref{table:obs_outflow_param}).
From the first moment (intensity weighted velocity) maps, characteristic velocities of the outflowing gas were derived.
These values are presented in the second column of Table~\ref{table:obs_outflow_param}. The uncertainties in
Table~\ref{table:obs_outflow_param} reflect the noise output from the ALMA observation simulator which were propagated
(in quadrature) through the equations used to derive the outflow properties.

Outflow momentum was calculated from $P=mv$, and energy from $E=0.5mv^2$. Outflow luminosities and mass loss rates
were determined by dividing the outflow energies and masses by the kinematic age of the outflows (400~yr).
More details on this type of calculation can be found in \citet{klaassenetal11}. For each of our modelled outflows,
we find that many outflow properties as measured by ALMA are about a factor of two higher than those observed with CARMA.

\begin{table*}
\caption{Outflow parameters derived from CARMA and ALMA simulations}
\begin{center}
\begin{tabular}{lrr@{$\pm$}lr@{$\pm$}lr@{$\pm$}lr@{$\pm$}lr@{$\pm$}lr@{$\pm$}lcc}
\hline\hline
CARMA&& \multicolumn{2}{c}{M} & \multicolumn{2}{c}{V}& \multicolumn{2}{c}{P}& \multicolumn{2}{c}{E} & \multicolumn{2}{c}{L}
& \multicolumn{2}{c}{$\dot{\rm M}$} & $R$\\
&& \multicolumn{2}{c}{($M_\odot$)} & \multicolumn{2}{c}{(km s$^{-1}$)} & \multicolumn{2}{c}{($M_\odot$ km s$^{-1}$)} & \multicolumn{2}{c}{(10$^{44}$ erg)}
& \multicolumn{2}{c}{($L_\odot$)} & \multicolumn{2}{c}{(10$^{-3}$ $M_\odot$ yr$^{-1}$)} & (AU)\\ \hline
Run A &
blue & 0.059 & 0.001 & 3.816 & 0.126 & 0.224 & 0.010 & 0.085 & 0.007 & 0.175 & 0.048 & 0.147 & 0.031 & 4100\\
&red & 0.066 & 0.001 & 3.391 & 0.150 & 0.223 & 0.012 & 0.075 & 0.007 & 0.155 & 0.046 & 0.165 & 0.035 & 4100\\
\hline
Run B (left)  &
blue & 0.013 & 0.000 & 3.634 & 0.199 & 0.048 & 0.004 & 0.017 & 0.002 & 0.036 & 0.012 & 0.033 & 0.007 & 3300\\
&red & 0.060 & 0.000 & 3.318 & 0.199 & 0.198 & 0.013 & 0.065 & 0.008 & 0.134 & 0.044 & 0.149 & 0.031 & 2100\\
\hline
Run B (right) &
blue & 0.058 & 0.000 & 3.549 & 0.140 & 0.206 & 0.009 & 0.073 & 0.006 & 0.150 & 0.043 & 0.145 & 0.030 & 5000\\
&red & 0.034 & 0.000 & 2.453 & 0.000 & 0.082 & 0.001 & 0.020 & 0.000 & 0.041 & 0.009 & 0.084 & 0.018 & 4100\\
\hline
\hline
ALMA&& \multicolumn{2}{c}{M} & \multicolumn{2}{c}{V}& \multicolumn{2}{c}{P}& \multicolumn{2}{c}{E} & \multicolumn{2}{c}{L}
& \multicolumn{2}{c}{$\dot{\rm M}$} & $R$\\
&& \multicolumn{2}{c}{($M_\odot$)} & \multicolumn{2}{c}{(km s$^{-1}$)} & \multicolumn{2}{c}{($M_\odot$ km s$^{-1}$)} & \multicolumn{2}{c}{(10$^{44}$ erg)}
& \multicolumn{2}{c}{($L_\odot$)} & \multicolumn{2}{c}{(10$^{-3}$ $M_\odot$ yr$^{-1}$)} & (AU)\\ \hline
Run A &
blue & 0.082 & 0.000 & 3.888 & 0.099 & 0.320 & 0.009 & 0.124 & 0.007 & 0.255 & 0.065 & 0.206 & 0.042 & 4100\\
&red & 0.101 & 0.000 & 3.297 & 0.154 & 0.333 & 0.016 & 0.109 & 0.010 & 0.225 & 0.066 & 0.252 & 0.051 & 4100\\
\hline
Run B (left) &
blue & 0.050 & 0.000 & 3.446 & 0.000 & 0.171 & 0.001 & 0.059 & 0.000 & 0.121 & 0.025 & 0.124 & 0.025 & 3300\\
&red & 0.141 & 0.000 & 2.757 & 0.184 & 0.388 & 0.026 & 0.106 & 0.014 & 0.219 & 0.073 & 0.352 & 0.071 & 2100\\
\hline
Run B (right) &
blue & 0.060 & 0.000 & 3.550 & 0.143 & 0.213 & 0.009 & 0.075 & 0.006 & 0.155 & 0.044 & 0.150 & 0.030 & 5000\\
&red & 0.044 & 0.000 & 2.370 & 0.000 & 0.104 & 0.000 & 0.024 & 0.000 & 0.050 & 0.010 & 0.109 & 0.022 & 4100\\
\hline
\hline
\end{tabular}
\tablecomments{The table shows the mass $M$, velocity $V$, momentum $P$, kinetic energy $E$, mechanical luminosity $L$, mass-loss rate $\dot{M}$,
and size $R$ of the observed outflows for two snapshots in Run~A and Run~B. The kinematic age of all outflows is $400\,$yr.}
\end{center}
\label{table:obs_outflow_param}
\end{table*}

\begin{figure*}
\vspace{10pt}
\begin{center}
\includegraphics[height=150pt]{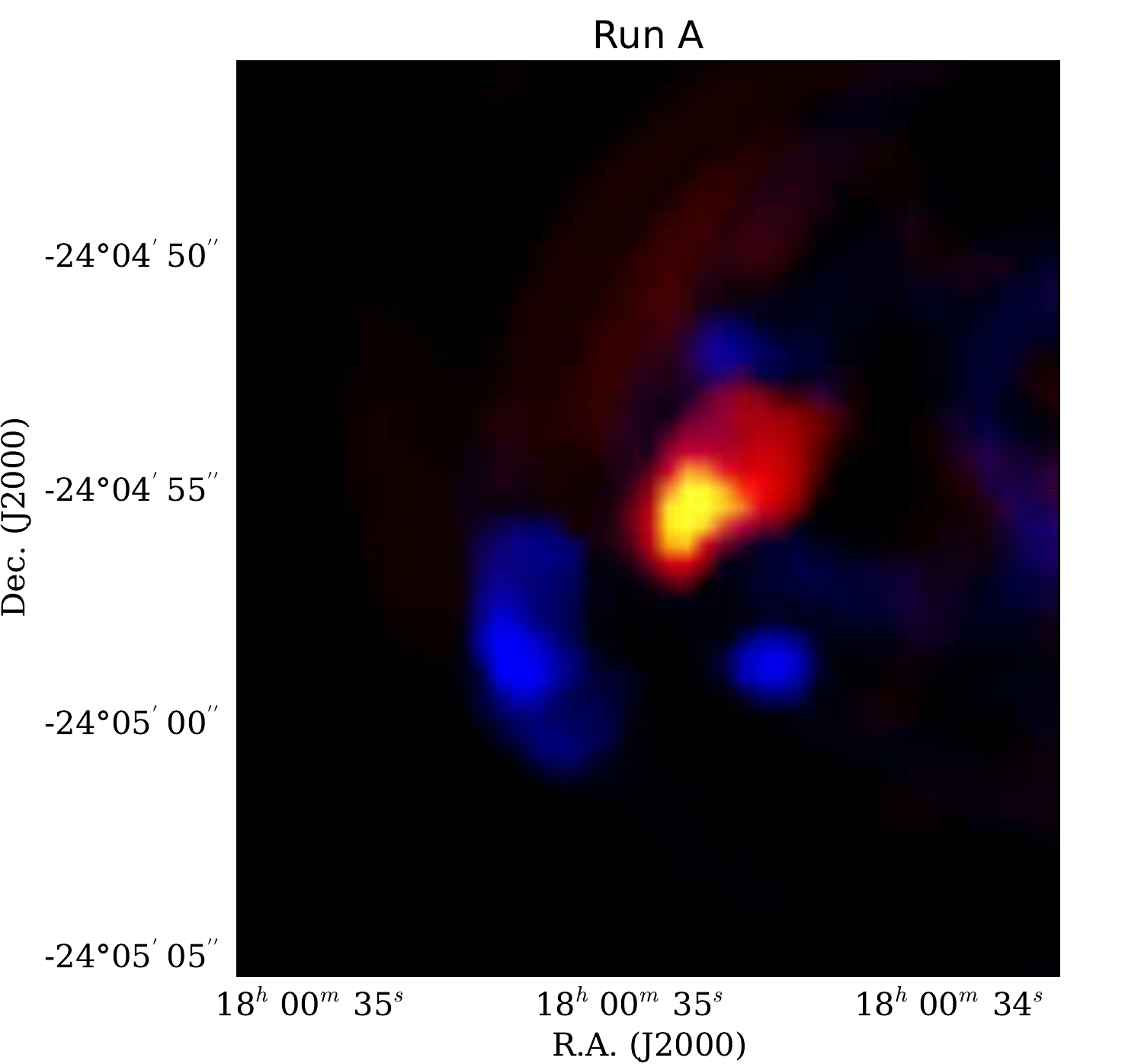}
\hspace{20pt}
\includegraphics[height=150pt]{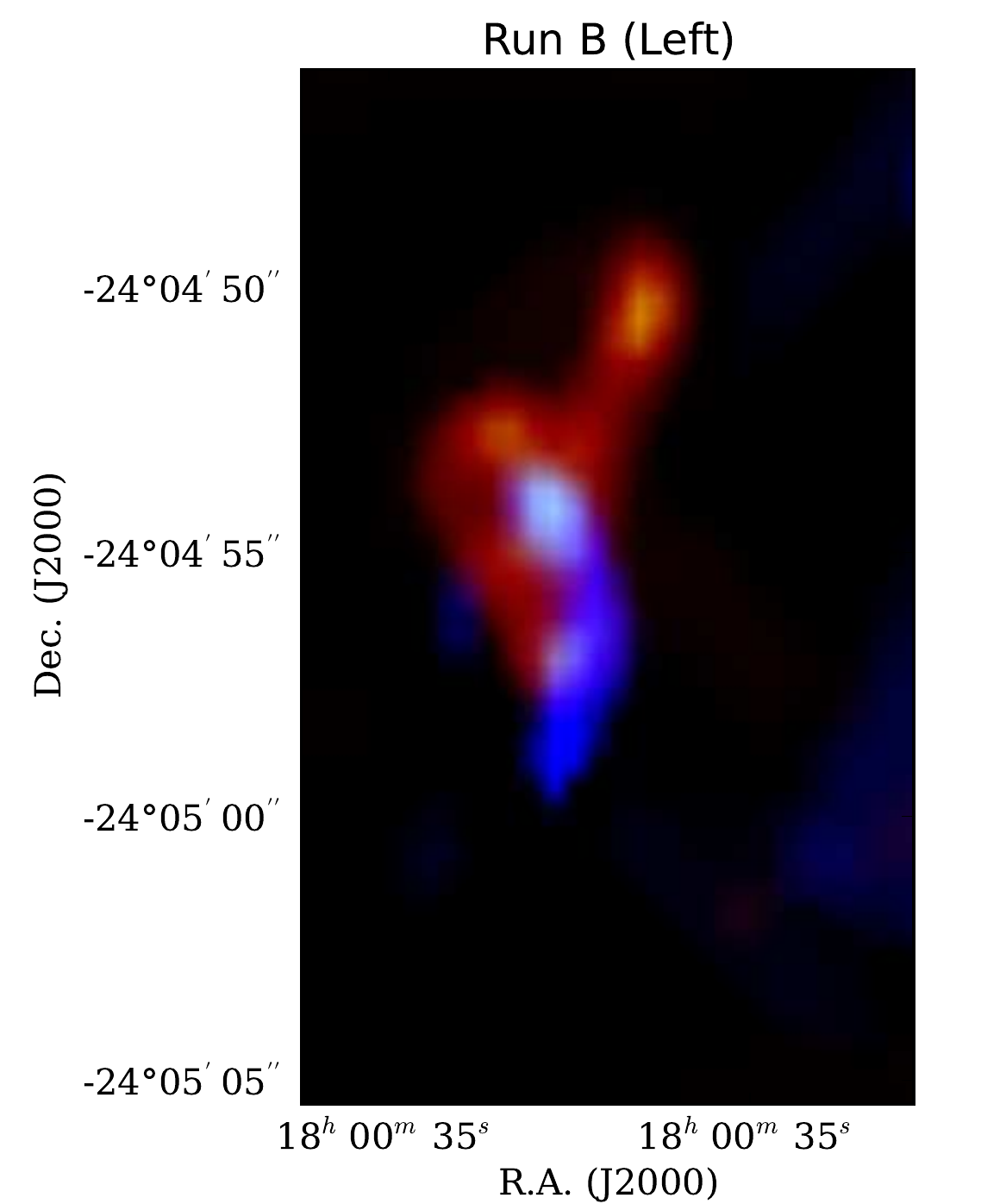}
\hspace{12pt}
\includegraphics[height=150pt]{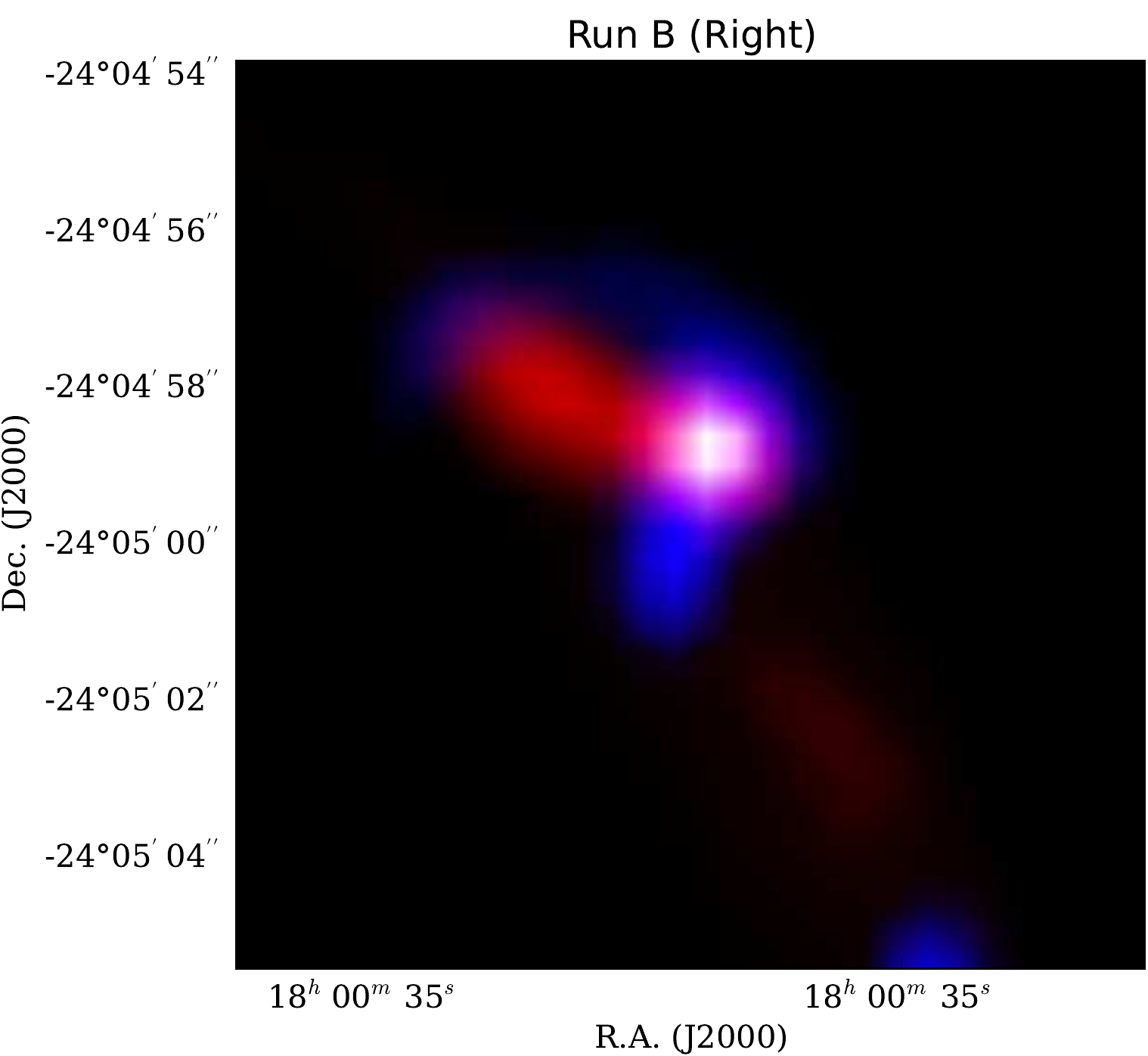}
\caption{Three color images of synthetic CARMA observations of snapshots from Run~A and Run~B processed with CASA. The resolution of these observations is
$1.3" \times 1.0"$, and the two point mosaic was simulated to have been observed for four hours on source. Noise
was added in the UV plane to simulate a system temperature of 200 K, and the source was assumed to be at
the distance of G5.89 (1.3~kpc).}
\label{fig:carma}
\end{center}
\end{figure*}

\begin{figure*}
\vspace{10pt}
\begin{center}
\includegraphics[height=150pt]{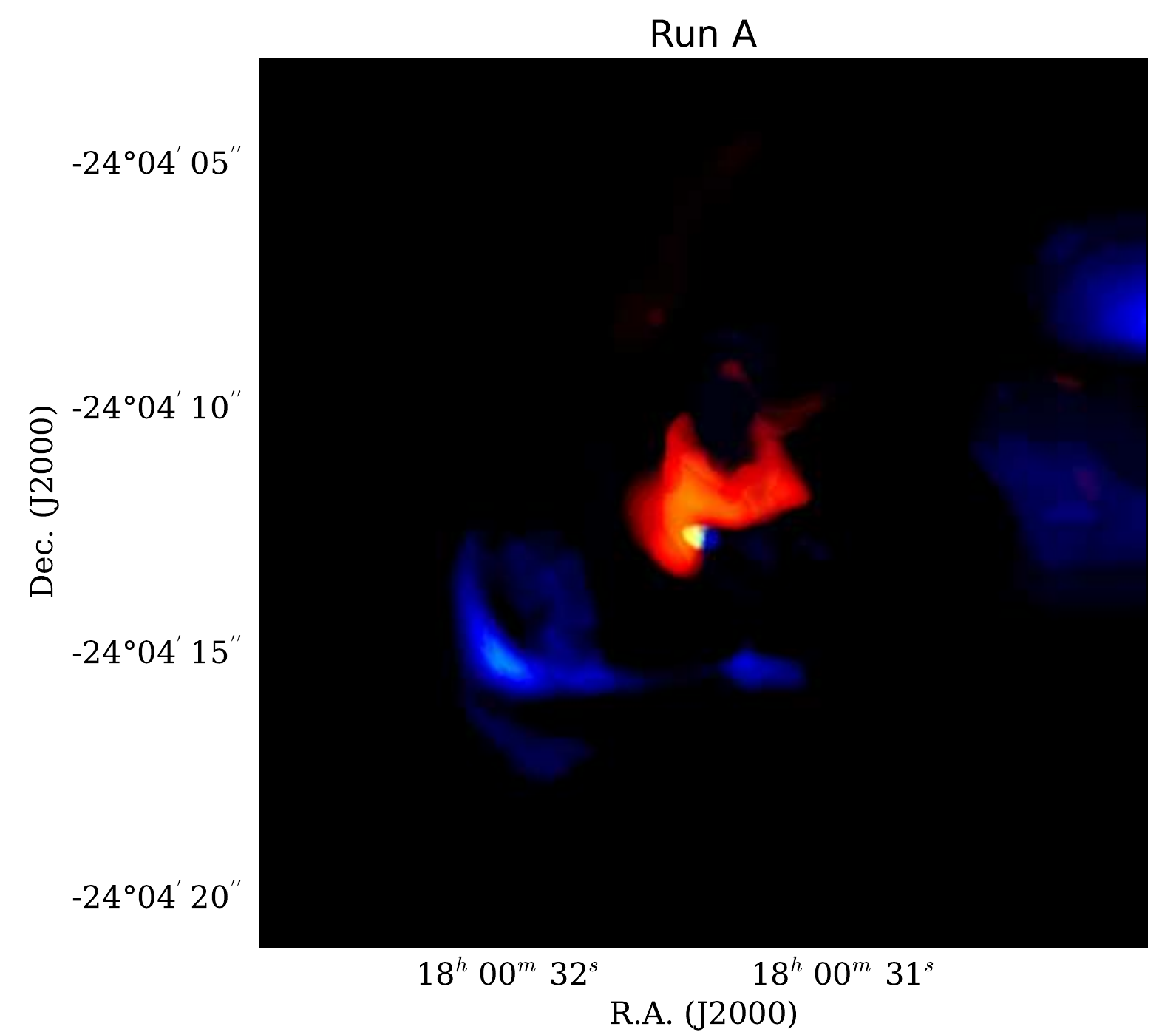}
\hspace{20pt}
\includegraphics[height=150pt]{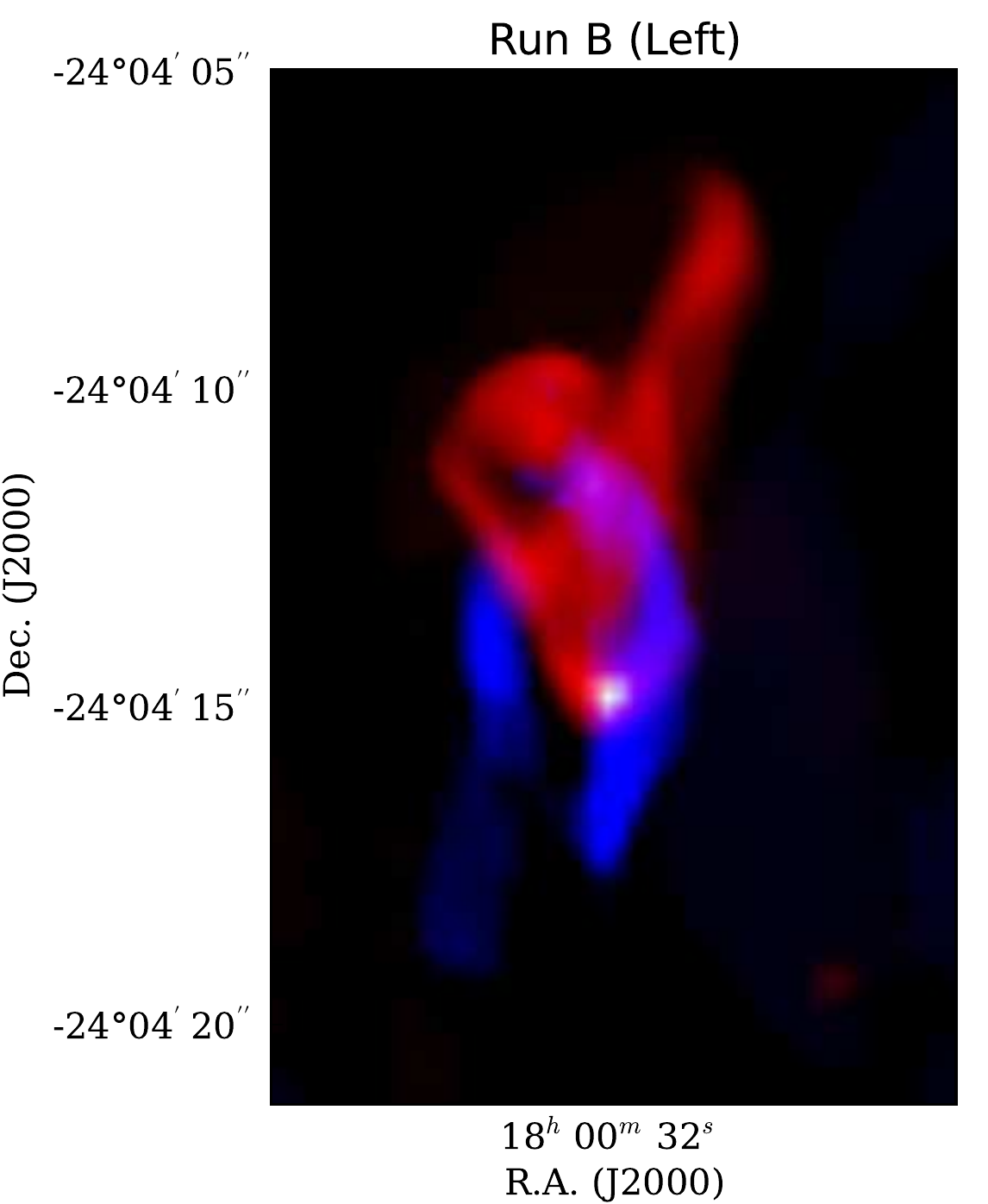}
\hspace{12pt}
\includegraphics[height=150pt]{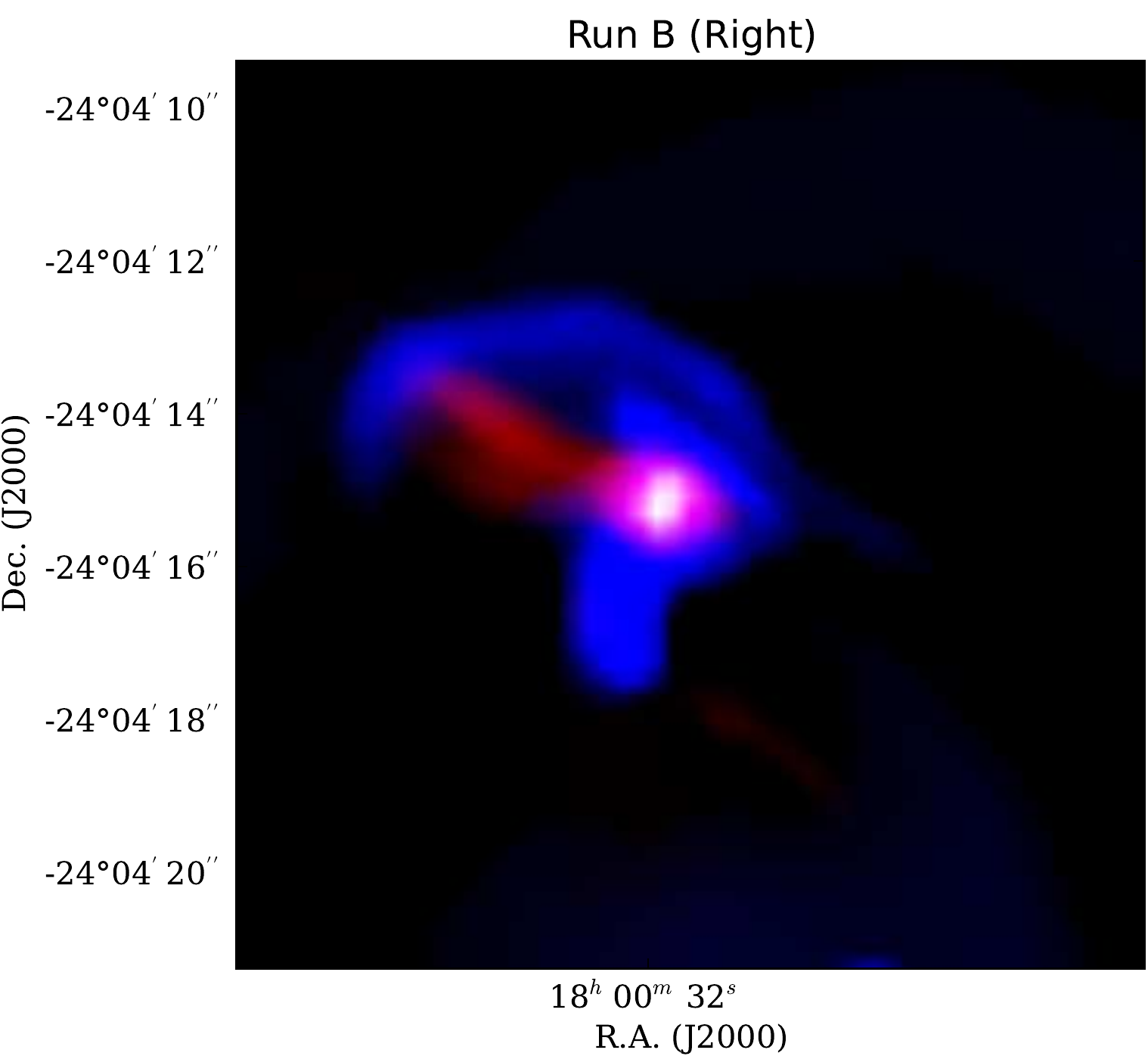}
\caption{Three color images of synthetic ALMA observations of snapshots from Run~A and Run~B processed with CASA. The resolution of these observations is
0.5$''$, and the six point mosaic was simulated to have been observed for four hours on source. Noise
was added in the UV plane to simulate fourth quartile observing conditions, and the source was assumed to be at
the distance of G5.89 (1.3~kpc).}
\label{fig:alma}
\end{center}
\end{figure*}

Figures~\ref{fig:carma} and~\ref{fig:alma} show three color images of the CO emission for snapshots of Run~A and Run~B as would be
observed with CARMA and ALMA, respectively, after being processed through CASA. The images were observed at an inclination
of 30$^\circ$ with respect to the disk normal direction, which is a typical orientation \citep{cabber86}.
The zeroth moment maps of the blue and red shifted emission were cut at 3~$\sigma$ for this plot
(as shown in blue and red, respectively). The white areas correspond to the brightest regions in the map,
the position of the high-mass protostar. The two outflows visible in the map for Run~B are driven by two separated massive stars, one of
which still has a small \hii\ region (right), while the \hii\ region around the other star has totally recombined (left).
As the stars move through the flattened structure, their ionizing radiation continuously drives material out perpendicular to it,
but after the star has moved and the gas has recombined, the gas falls back. This process creates a large-scale
molecular fountain, but its velocity is smaller than our low-velocity cut-off, and the interferometer additionally filters
out some of its large-scale structure. Hence, the map only shows the smaller, more powerful outflows that are currently
driven by the ionization feedback from the massive stars, and not the large-scale bubble that is also present.

\section{Discussion}
\label{sec:disc}

\subsection{Comparison to exact solution}

We can calculate the mass, momentum and kinetic energy in the outflows exactly. To this end, we
use RADMC-3D to sum over the mass, momentum and energy in zones with velocity high enough to be contained in the outflow as viewed along the
line of sight of the simulated observation.
We only take cells with $3\,$km/s\,$ \leq v \leq 7\,$km/s into account, which corresponds to the selection criterion used with the
synthetic maps. The results of these direct measurements are listed in Table~\ref{table:direct}. 
The directly measured values are comparable with the numbers inferred from the synthetic observations, but generally higher.
The reason could be that the synthetic observations do not recover all the flux from the outflows.

The morphology of the outflow for Run~A as pictured in Figure~\ref{fig:alma} suggest that the outflow is pointing mostly along
the line of sight to the observer, which can be confirmed by examining the raw simulation data. This finding is also supported by the
intensity-weighted velocities (characteristic velocities, listed in
Table~\ref{table:obs_outflow_param}) of the gas in this outflow, which are slightly higher than those for the outflows in Run~B.
The morphology of the two outflows in the figure for Run~B (along with their lower characteristic velocities) suggest
they are more likely oriented closer to the plane of the sky, which is again confirmed by the raw data.
However, the ionizing luminosity that powered these outflows before the \hii\ region recombined is also vastly different: the star in Run~A has
57~$M_\odot$, while the two massive stars in Run~B have masses around 20~$M_\odot$ each.

Our analysis shows that the flickering of the ionization feedback can produce molecular outflows that are morphologically similar to but smaller than the observed
ones. They appear to be very young compared to the age of the star because they are not driven continuously. Because of their short
kinematic age, the momentum, energy and luminosity of the outflows are relatively small compared to observed outflows around high-mass stars. We predict that these small, high-momentum,
but low-velocity outflows will be found by upcoming ALMA observations that probe the driving sources of high-mass outflows, if they can be disentangled
from the larger high-velocity outflows.

\subsection{Comparison to observations}

The outflows in Figure~\ref{fig:alma} are smaller than those generally seen in high-mass star forming regions.
However their kinematic ages ($\sim$ 400 yr) suggest they are also much younger than those which have been observed to date.
These two properties are presumably correlated. Although the calculated outflow masses and kinematic properties (momentum, energy and mechanical luminosity) are consistent with
the range of these properties found for high-mass star-forming regions \citep[e.g.][]{wuetal04}, they are located near their lower boundaries.

We find the outflow velocities in the simulations to be an order of magnitude smaller than those in G5.89 even though previous observations of this source
\citep[e.g.][]{pugaetal06,suetal12} were at similar angular resolutions.

The simulated mass infall rates are high compared to observations if we assume all of the infalling mass makes it onto the central forming star
($\dot{M}_\mathrm{in} \sim 10^{-3}$ $M_\odot$ yr$^{-1}$ suggests a 10 $M_\odot$ star forms in 10$^4$ yr), however much of the infalling material will
not be accreted by the central star(s). Instead, secondary stars form in the gravitationally
unstable accretion flow around the massive star and intercept gas that would otherwise fall inwards in a process called fragmentation-induced starvation
\citep{petersetal10a,petersetal10c,girietal12}.

\begin{table}
\caption{Outflow parameters derived from direct measurements}
\begin{center}
\begin{tabular}{lrccc}
\hline\hline
&& \multicolumn{1}{c}{M} & \multicolumn{1}{c}{P}& \multicolumn{1}{c}{E} \\
&& \multicolumn{1}{c}{($M_\odot$)} & \multicolumn{1}{c}{($M_\odot$ km s$^{-1}$)} & \multicolumn{1}{c}{(10$^{44}$ erg)}\\ \hline
Run A &
blue&0.67&2.43&0.91\\
&red&1.49&5.84&2.44\\
\hline
Run B (left)  &
blue&0.04&0.18& 0.08\\
&red&0.23&0.90&0.38\\
\hline
Run B (right) &
blue&0.13&0.53&0.23\\
&red&0.04&0.17&0.07\\
\hline
\hline
\end{tabular}
\tablecomments{Outflow mass $M$, momentum $P$ and kinetic energy $E$ as directly measured from the simulation data.}
\end{center}
\label{table:direct}
\end{table}

\section{Conclusions}
\label{sec:conc}

Our results show that ionization feedback is able to drive molecular outflows around high-mass stars that have properties consistent with some observed ones.
However, the measured properties are rather on the lower boundary of the typical values of high-mass outflows, which makes it unlikely that ionization feedback
is the main driver of high-mass outflows. Additional feedback mechanisms like radiation pressure and stellar winds or magnetic fields might help,
but as discussed in the Introduction, there is no strong evidence that this is indeed the case.

We note, however, that lower-mass companions naturally form with high-mass stars in their gravitationally
unstable accretion flow \citep{petersetal10a,petersetal10c,petersetal11a,girietal12}, in agreement with the observation that massive stars
are typically found as members of higher-order multiple stellar systems and clusters \citep{hohaschik81,zinnyork07}.
We propose that it is the joint outflow activity of the low- and intermediate-mass companions that we are seeing.

\begin{figure}
\begin{center}
\includegraphics[width=8cm]{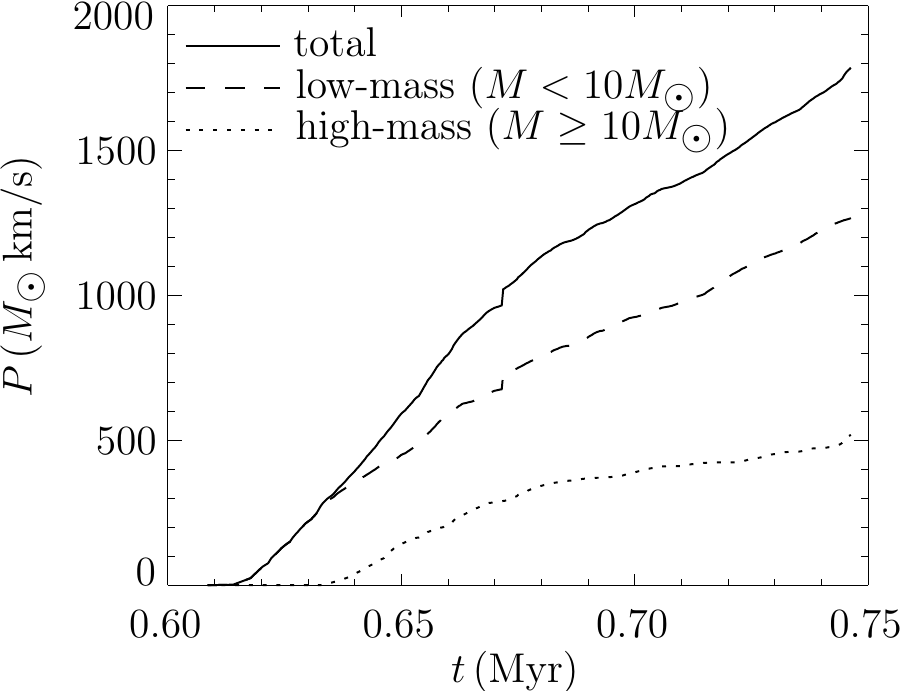}
\caption{Momentum input $P$ according to the outflow model from the stars in the cluster
as function of time $t$. The joint momentum injected by the low- and intermediate-mass stars is well within
the regime of high-mass outflows.}
\label{fig:lowmass}
\end{center}
\end{figure}

We tested whether the low- and intermediate-mass companions in the accretion flow around the massive stars in our simulations at least in principle are able
to reproduce the characteristics of high-mass outflows with a simple model. We assume that
a factor of $f=0.1$ of the accreted material onto the stars is ejected
again as an outflow with an initial velocity of $v_\mathrm{wind} = 150\,$km\,s$^{-1}$ \citep[e.g.][]{bontetal96}. The momentum
injected into the surrounding gas by a single star per unit time is then $\dot{P} = f\, v_\mathrm{wind}\, M_\mathrm{acc}$
for an accretion rate $M_\mathrm{acc}$. Figure~\ref{fig:lowmass} shows the integrated total momentum $P$
as function of time for all stars in the cluster.

The figure shows that even with conservative assumptions the low- and intermediate-mass stars alone can easily produce outflow momenta
in the regime of observed high-mass outflows. Their contribution is orders of magnitude larger
than the momentum in the ionization-driven outflows.

Further simulations that simultaneously include ionization feedback from massive stars as well as the
mechanical feedback from their low- and intermediate-mass companions
are required to thoroughly test this proposal.

\section*{Acknowledgements}

We thank Herwig Zilken from the J\"ulich Supercomputing Centre for his help with visualizing our simulation data.
We also thank the anonymous referee for useful comments that helped to improve the paper.
T.P. acknowledges financial support as a Fellow of the Baden-W\"{u}rttemberg Stiftung funded
through contract research under grant P-LS-SPII/18 in their program {\em Internationale Spitzenforschung II} and through SNF grant
200020\textunderscore 137896.
M.-M.M.L. was partly supported by NSF grant AST 11-09395.
R.S.K. acknowledges funding from the {\em Deutsche Forschungsgemeinschaft} DFG via grants KL~1358/14-1 as part of
the Priority Program SPP 1573 {\em Physics of the Interstellar Medium} as well as via the Collaborative Research Project
SBB~811 {\em The Milky Way System} in subprojects B1, B2, and B4.
R.B. acknowledges funding from the DFG via the grant BA 3706/1-1.
We acknowledge computing time at the Leibniz-Rechenzentrum in Garching, the
Texas Advanced Computing Center through grant TG-MCA99S024 from the Extreme Science and Engineering Discovery Environment (XSEDE), which
is supported by the National Science Foundation grant number OCI-1053575, and at J\"ulich Supercomputing Centre. 
The FLASH code was in part developed by the DOE-supported Alliances Center for Astrophysical Thermonuclear Flashes (ASCI) at the University of Chicago.

\end{document}